\documentclass[aps,prb,twocolumn,groupedaddress,showpacs,showkeys,floatfix,10pt]{revtex4-1}

\usepackage{graphicx}
\usepackage{epsfig}

\usepackage{amssymb}
\usepackage{amsmath}
\usepackage{amsthm}
\usepackage{color}

\usepackage{sidecap}
\sidecaptionvpos{figure}{c}

\begin{document}

\title{Strong Amplitude and Phase Modulation of \\Optical Spatial Coherence with Surface Plasmon Polaritons}

\author{Dongfang Li}
\author{Domenico Pacifici}
\email{Domenico\_Pacifici@brown.edu}
\affiliation{School of Engineering, Brown University, Providence, RI 02912, USA}

\date{\today}

\begin{abstract}
The degree of optical spatial coherence---a fundamental property of light that describes the mutual correlations between fluctuating electromagnetic fields---has proven challenging to control at the micrometer scale. Here we employ surface plasmon polaritons---evanescent waves excited on both surfaces of a thin metal film---as a means to entangle the random fluctuations of the incident electromagnetic fields at the slit locations of a Young's double-slit interferometer. Strong tunability of the complex degree of spatial coherence of light is achieved by finely varying the separation distance between the two slits. Continuous modulation of the degree of spatial coherence with amplitudes ranging from 0\% up to 80\% allows us to transform totally incoherent incident light into highly coherent light, and vice versa. These findings pave the way for alternative methods to engineer flat optical elements with multi-functional capabilities beyond conventional refractive- and diffractive-based photonic metasurfaces.
\end{abstract}

\maketitle
Control of light propagation and interaction with matter relies upon knowledge of the amplitude and phase of the electromagnetic fields, as well as their mutual temporal and spatial correlations, which are described by the complex degree of coherence~\cite{mandel1995optical,wolf2007introduction,Wolf_PLA}. Planar metal and dielectric surfaces with subwavelength features---also known as metasurfaces \cite{kildishev2013planar,yu2014flat,lin2014dielectric}---have recently been proposed as a way to manipulate light by producing abrupt changes in the local amplitude and phase of the incident electromagnetic fields \cite{yu2011light}. Typically, the incident fields are highly coherent and the scattered fields originating from the interaction of incident light with the nano-structured surfaces are mutually coherent. Then, optical interference effects produce desired changes in beam directionality, polarization, intensity, phase, and spin~\cite{aieta2015multiwavelength,maguid2016photonic}.

Optical spatial coherence can provide for an alternative, powerful tool to control the flow of light with various degrees of coherence beyond conventional wavefront shaping methods offered by photonic metasurfaces. Indeed, from an application standpoint, control of the  degree of spatial coherence of light can lead to higher-resolution speckle-free imaging~\cite{Redding_NatPhoton,Knitter}, ultimate capability to shape~\cite{kip2000modulation} and redirect light beams~\cite{Dijk_PRL,Wang_PRA_2015,Wang_PRA_2016}, 
and enable novel modulation methods for free-space optical communications and interconnects~\cite{anderson1995spatial,Gbur_JOSAA}. However, full control of spatial coherence at length scales comparable to the wavelength of light has proven challenging. Although it has theoretically been suggested that surface plasmon polaritons (SPPs)---electromagnetic waves evanescently bound to metal surfaces---can in principle modulate the degree of spatial coherence~\cite{Gan_PRL,Gan_Plasm}, a very few experimental studies have shown this effect, reporting only modest modulation amplitudes~\cite{Kuzmin,Ravets,Divitt_OpLett}. 
 
Here we show, for the first time, extremely large, finely tunable changes in the amplitude and phase of the complex degree of spatial coherence, controllable at the micrometer scale by simply varying the slit-slit separation distance, as well as the wavelength and polarization state of the incident light. For instance, we report how light incident on a Young's double-slit can be tuned from completely incoherent to partially coherent, and vice versa, with degrees of coherence continuously variable from $\sim$0\% (totally incoherent) to $\sim$80\% (almost fully coherent).  

\begin{figure*}[h]	
  \centering
  \includegraphics[width=1\textwidth]{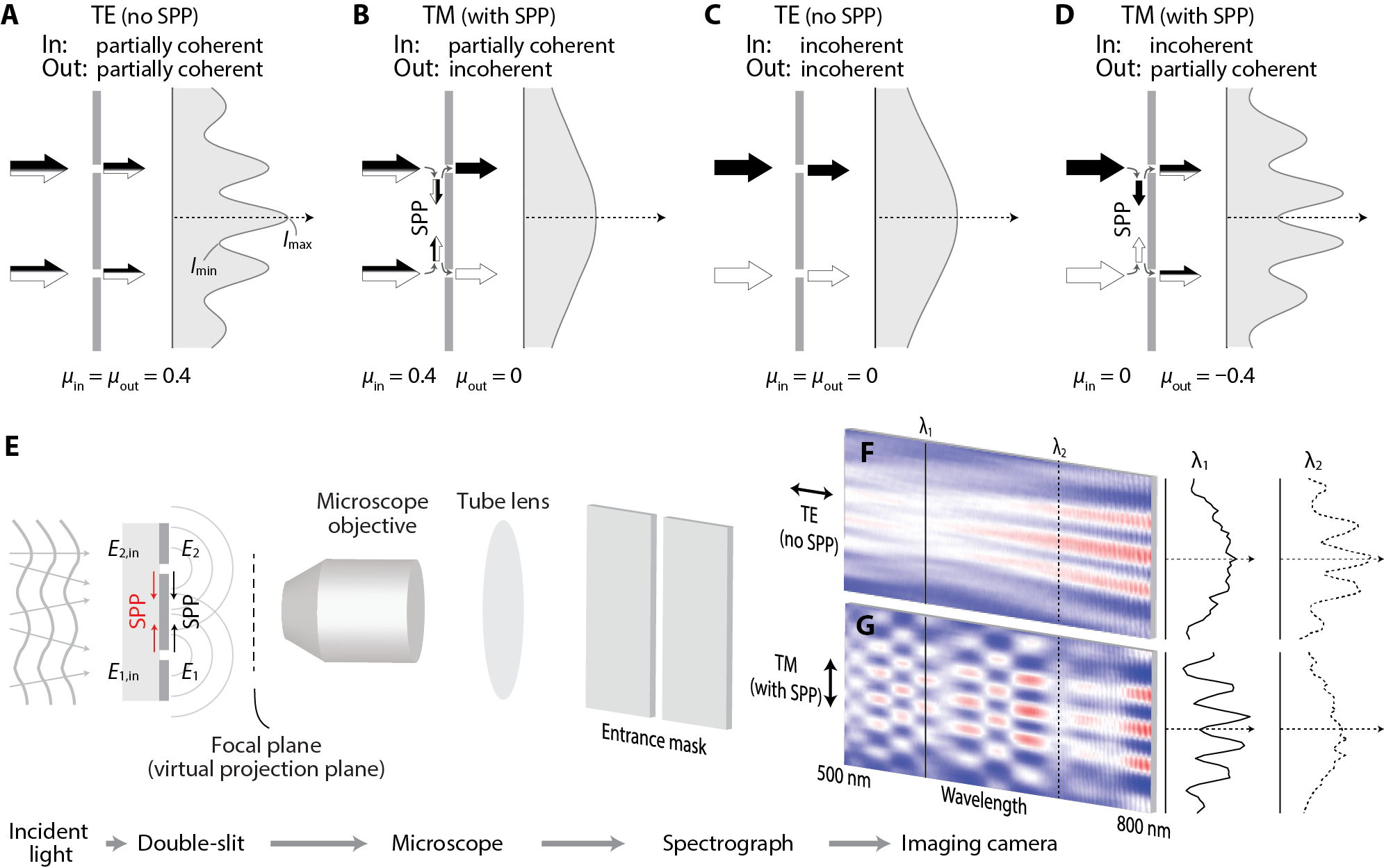}
    \caption{\textbf{Schematics of Young's double-slit experiment with and without surface plasmon polaritons.} Simulated interference patterns showing modulation of the degree of spatial coherence of incident light induced by SPPs~\cite{SuppMat}. Partially coherent light (\textbf{A}) can be transformed into incoherent light (\textbf{B}), and, vice versa, incoherent light (\textbf{C}) can be transformed into partially coherent light (\textbf{D}) upon interaction with the slitted screen, when SPPs are excited. Electromagnetic fields with different degrees of spatial coherence incident on the two slits are schematically indicated with arrows filled with different fractions of black and white areas. (\textbf{E}) Schematic of the experimental setup designed to record the far-zone interference pattern from Young's double-slit to extract visibility as a function of wavelength. (\textbf{F}) Experimental TE (no SPPs) and (\textbf{G}) TM (with SPPs) wavelength-resolved far-zone interference patterns originating from a Young's double-slit with slit-slit separation distance $d=5$~$\mu$m and subtended illumination angle $\Delta \theta \approx$~6$^\circ$. \textbf{Insets to F, G (right panels):} experimental SPP-induced modulation of interference patterns at two different wavelengths, showing transformation of incoherent light into partially coherent light (solid lines, $\lambda_\mathrm{1}=581$~nm), and vice versa (dashed lines, $\lambda_\mathrm{2}=712$~nm), when the polarization state of the incident light is varied from TE to TM.
}
\label{figSetup}
\end{figure*}

\paragraph*{\textbf{Physical mechanism enabling optical coherence modulation with surface plasmons.}}
Partially coherent light exiting two slits placed at points $P_1$ and $P_2$ in an opaque film can generate interference fringes when projected onto a far-zone screen, as the result of constructive and destructive interference caused by the difference in the optical paths from each slit to a specific point on the screen (Fig.~\ref{figSetup}A and fig.~S1). The fringe contrast, or \emph{visibility}, defined as $\mathcal{V}=(I_\mathrm{max}-I_\mathrm{min})/(I_\mathrm{max}+I_\mathrm{min})$, where $I_\mathrm{max}$ and $I_\mathrm{min}$ are the maximum and minimum intensities in the interference pattern (Fig.~\ref{figSetup}A), can be used to quantitatively describe the complex mutual correlation function $\mu(P_1,P_2)=|\mu|e^{i\phi}$ between the electromagnetic fields at the two different points where the slits are located. Indeed, it can be proven that the visibility of the interference fringes is equal to the amplitude of the degree of spatial coherence, that is $|\mu|=\mathcal{V}$~\cite{Mandel_JOSA}. The phase $\phi$  can be inferred by looking at the interference condition at the center of the screen, with constructive (destructive) interference corresponding to $\phi=0$ ($\pi$). Therefore, Young's double-slits can be employed to directly measure and quantify the changes of the degree of coherence of incident light at the slit locations, with $\mu=0$~(or $1$) corresponding to totally incoherent (or coherent) light.

\paragraph*{\textbf{Surface plasmons transform incoherent light into coherent light and vice versa.}}
In the absence of SPPs, the degree of coherence of transmitted light is unaffected after interaction with the slitted screen (Fig.~\ref{figSetup}A), resulting in $\mu_\mathrm{in}=\mu_\mathrm{out}$. This condition can be realized by using perfectly opaque screens, or very large slit-slit separation distances in real metals, or when light is linearly polarized with the electric field direction parallel to the long axis of the slits (transverse-electric, or TE-polarized).

Under TM illumination, SPPs are excited and can introduce additional or eliminate existing correlations between the fields at the two slits, which enables strong modulation of the degree of spatial coherence provided that the SPP excitation process is  coherent~\cite{Altewischer,Tame,Wang_OpLett}. For example, Fig.~\ref{figSetup}B shows how SPPs can transform partially coherent light incident onto the slits into fully incoherent light leaving the slits. This is evidenced by the disappearance of the interference fringes that are replaced by a broad, featureless  intensity distribution on the projection screen, caused by the incoherent sum of two single-slit diffraction patterns (Fig.~\ref{figSetup}B)~\cite{SuppMat}. 

Totally incoherent light incident upon the double slits is unable to generate interference fringes in the far field (Fig.~\ref{figSetup}C, TE-polarization, no SPPs). However, SPPs excited at each slit location and propagating toward the other slit can, under proper conditions, mix the initially uncorrelated fields and generate partially coherent light with non-zero degree of coherence at the output of the double-slit. Thanks to these SPP-induced correlations introduced by the alternative path that light can take in the form of propagating SPPs, totally incoherent light can be transformed into partially coherent light, and as a result the far-zone fringe visibility is restored (Fig.~\ref{figSetup}D). Far-zone interference patterns can therefore be used to infer the complex degree of spatial coherence and quantify the changes induced by SPPs.

\paragraph*{\textbf{Experiment.}}
Figure~\ref{figSetup}E shows a schematic of the experimental setup that was specifically designed to measure wavelength-resolved, Young's double-slit interference patterns, as a function of slit-slit separation distance $d$, and incident polarization. In particular, double-slits etched in a thin silver film were illuminated by linearly-polarized broadband light under K\"ohler illumination~\cite{Morrill}. The objective of an inverted microscope was purposely defocused to project the interference pattern onto the entrance mask of a spectrograph, which disperses the transmitted light and projects the generated interference patterns onto a two-dimensional imaging camera. Representative wavelength-resolved interference patterns emerging from a Young's double-slit with $d=5$~$\mu$m, illuminated with partially coherent light with two different polarization states (TE, Fig.~\ref{figSetup}F, and TM, Fig.~\ref{figSetup}G), clearly show the dramatic changes in fringe visibility caused by SPPs. Remarkably, the results reported in the right insets to Fig.~\ref{figSetup}, F and G, show---now experimentally---that SPPs can indeed transform incoherent light incident upon the double-slit into partially coherent light at the output of the slitted screen (solid lines,  $\lambda_{1}=581$~nm), as attested by the increased fringe visibility when TM-polarized light  is employed. Vice versa, partially coherent light can be transformed into incoherent light (dashed lines, $\lambda_\mathrm{2}=712$~nm) as evidenced by the suppressed fringe contrast due to SPP-induced modulation of the mutual correlation function.

\begin{figure*}[h]	
  \centering
  \includegraphics[width=1\textwidth]{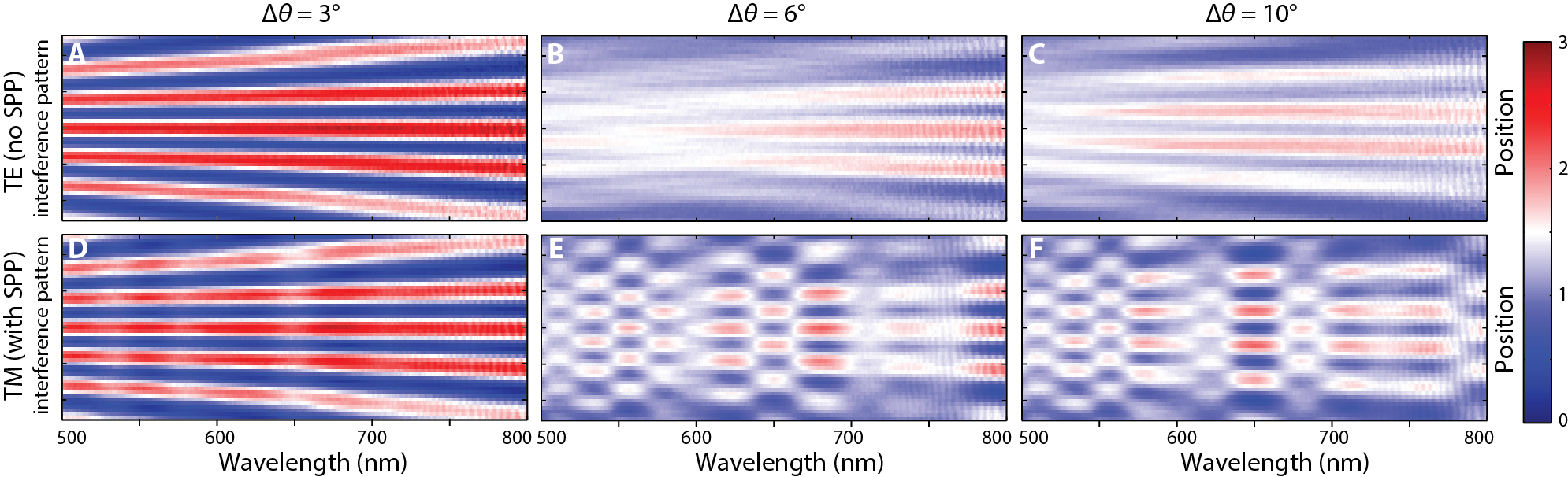}
	\caption{\textbf{Effects of surface plasmon polaritons on wavelength-resolved far-zone interference patterns for incident light with varying degrees of spatial coherence.}  Interference patterns measured at the output of a Young's double-slit interferometer with \textit{d}~=~5~$\mu$m as a function of incident wavelength, for three values of K\"ohler subtended illumination angles, $\Delta \theta \approx$~3$^\circ$, 6$^\circ$, 10$^\circ$, corresponding to decreasing spatial coherence, in the absence (TE-polarization, \textbf{A}--\textbf{C}) and in the presence (TM-polarization, \textbf{D}--\textbf{F}) of SPPs. Strong modulation of fringe visibility induced by SPPs is clearly visible.}
    \label{figFringe}
\end{figure*}

\paragraph*{\textbf{Effects of spatial coherence of incident light.}}
Several wavelength-resolved interference patterns for the same Young's double-slit with $d=5$~$\mu$m  under varying degrees of spatial coherence of the incident light are reported in Fig.~\ref{figFringe}. Under TE-polarized illumination (no SPPs), the far-zone interference patterns  gradually change as a function of incident wavelength $\lambda$, at any given $\Delta \theta$ (Fig.~\ref{figFringe}, A to C). Light with longer wavelength generates interference fringes that are spatially more spread out compared with shorter wavelengths, as the result of the longer optical path difference required to achieve the same phase shift on the projection screen. The fringe visibility is greatly reduced when light with lower degree of spatial coherence (that is, higher subtended illumination angle $\Delta \theta$) is used. By rotating the polarization state of the incident light by $\pi/2$ (that is, from TE to TM), SPPs from both metal/dielectric interfaces can be turned on and strong modulation of the fringe visibility is observed as the result of additional beatings between the two SPP modes (Fig.~\ref{figFringe}, D to F). Indeed, both amplitude (related to fringe visibility) and phase (related to either constructive or destructive interference at the screen center) of the degree of spatial coherence are strongly modulated under TM-polarized illumination, due to SPP excitation (see Fig.~\ref{figFringe}, E and F) that determines a redistribution of the far-zone light intensity. Both enhancement and suppression of fringe visibility can be achieved relative to the respective interference patterns under TE illumination (no SPPs, Fig.~\ref{figFringe}, B and C), 

\paragraph*{\textbf{Theory of optical coherence modulation with surface plasmons.}}
An analytical expression for the output fields that includes SPP contributions can be written as $E_\mathrm{i}(\lambda,d)=\tau E_\mathrm{i,in}(\lambda)+\tau \beta E_\mathrm{j,in}(\lambda)$, where $(i,j)=(1,2)$ or $(2,1)$, $E_\mathrm{i,in}$ is the incident scalar field at one slit location, $\tau$ is the transmission coefficient through the slit, $\lambda$ is the free-space wavelength of light, $\beta=\beta_\mathrm{t}e^{ik_\mathrm{SPP,t}d}+\beta_\mathrm{b}e^{ik_\mathrm{SPP,b}d}$ with $\beta_\mathrm{t}$ and $\beta_\mathrm{b}$ the SPP coupling coefficients at the top (i.e., glass/Ti/Ag) and bottom (i.e., Ag/air) interfaces, and $k_\mathrm{SPP,t}$, $k_\mathrm{SPP,b}$ are the corresponding SPP complex wavevectors, that account for SPP absorption due to ohmic losses in the metal~\cite{SuppMat}. 

Using the definition of cross-spectral density function $W_\mathrm{ij}(\lambda)=\langle E^*_\mathrm{i}(\lambda)E_\mathrm{j}(\lambda)\rangle$, where the angle brackets indicate ensemble averaging, the degree of spatial coherence of light exiting the double-slit can be written as~\cite{Gan_PRL,Divitt_OpLett}:
\begin{equation*}
  \mu_\mathrm{out}(\lambda)=\frac{W_\mathrm{12}(\lambda)}{\sqrt{W_\mathrm{11}(\lambda)W_\mathrm{22}(\lambda)}}=
\end{equation*}
\begin{equation}
  =\frac{\mu_\mathrm{in}(\lambda)+2\mathrm{Re}(\beta)+|\beta|^2\mu_\mathrm{in}^{*}(\lambda)}{\sqrt{(1+|\beta|^2+2\mathrm{Re}[\beta\mu_\mathrm{in}(\lambda)])(1+|\beta|^2+2\mathrm{Re}[\beta\mu_\mathrm{in}^{*}(\lambda)])}},
  \label{eqmu12}
\end{equation}
where $\mu_\mathrm{in}$ is the complex degree of spatial coherence of light incident on the double-slit.

\begin{figure*}[h]	
\centering
\includegraphics[width=1
\textwidth]{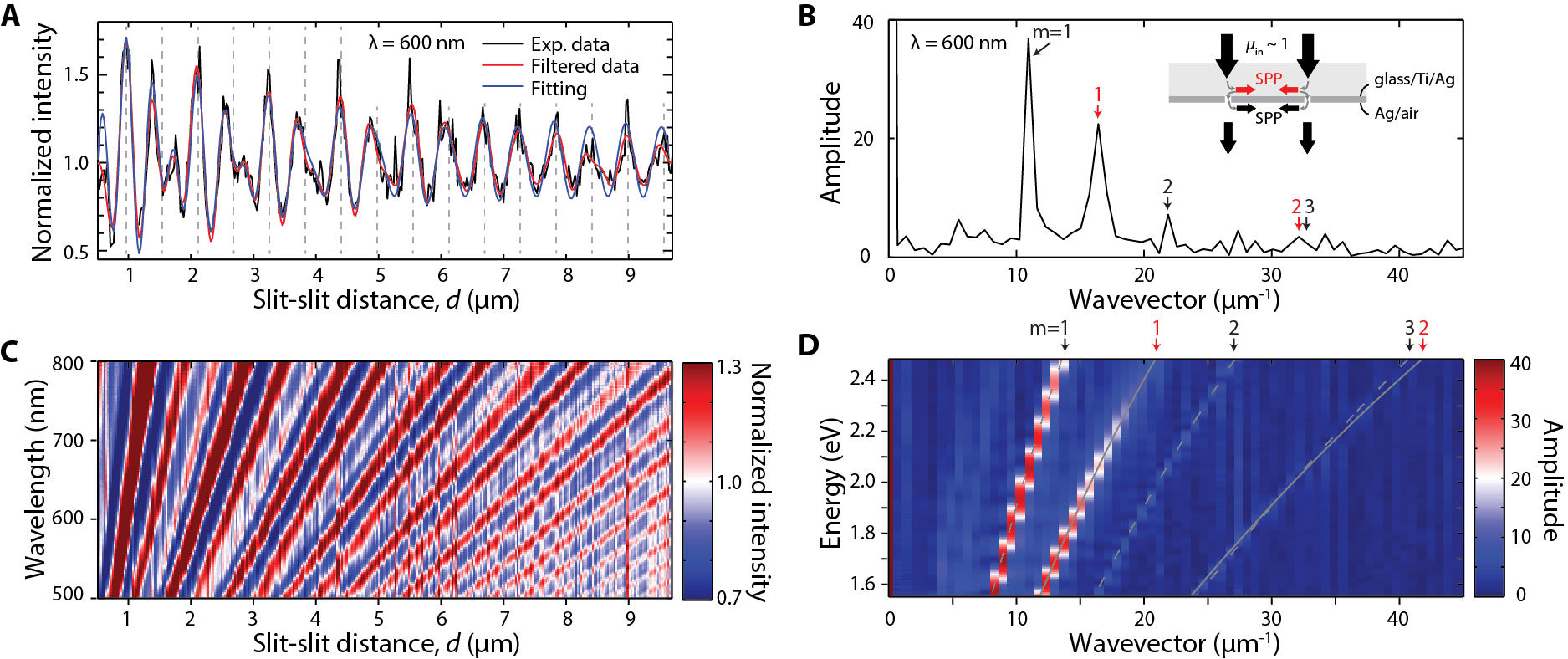}
	\caption{\textbf{Evidencing SPP contributions supported by both metal/dielectric interfaces.} (\textbf{A}) Plasmonic interferogram as a function of slit-slit distance measured at $\lambda=$~600~nm (black line). Reconstructed filtered data (red line) and theoretical fits (blue line) are also reported~\cite{SuppMat}. The distance between adjacent vertical dashed lines is equal to $\lambda_\mathrm{SPP,b}$, that is the SPP wavelength at the Ag/air interface. (\textbf{B}) Discrete Fourier transform power spectrum calculated from the experimental data in \textbf{A}, which clearly shows different orders of SPP contributions from both glass/Ti(3 nm)/Ag (red arrows) and Ag/air (black arrows) interfaces. \textbf{Inset:} schematic of SPPs propagating along both metal/dielectric interfaces and simultaneously affecting the output intensity and far-zone interference pattern. (\textbf{C}) Wavelength-resolved plasmonic interferograms obtained by normalizing the transmission spectra through double-slit interferometers by the reference transmission spectra through individual slits. (\textbf{D}) Energy-resolved power spectra obtained by applying discrete Fourier transform to the data in \textbf{C}. Energy-momentum dispersion curves evidence the presence of SPPs supported by both Ag/dielectric interfaces (bright bands in \textbf{D}), and are in good agreement with calculations (gray lines) based on the dielectric functions of materials~\cite{SuppMat}.}
    \label{figFFT}
\end{figure*}

\paragraph*{\textbf{Beatings between surface plasmons excited on both metal/dielectric interfaces.}}
To better quantify and control the modulation of the complex degree of spatial coherence, SPP contributions from both metal/dielectric interfaces need to be considered and carefully tailored. These contributions can be extracted by analyzing the plasmonic interferogram, that is a plot of the normalized intensity ratio at a given wavelength obtained by normalizing the light intensity transmitted through each double-slit with varying separation distance to the reference intensity transmitted through the corresponding individual slit~\cite{SuppMat}. Figure~\ref{figFFT}A shows a representative plasmonic interferogram measured at $\lambda=$~600~nm. Discrete Fourier transform analysis applied to this data set~\cite{Morrill,Li_OE} clearly shows the different SPP contributions originating from both Ag/air (black arrows) and glass/Ti/Ag interfaces (red arrows) as distinct intensity peaks in the Fourier power spectrum amplitude (Fig.~\ref{figFFT}B). This procedure can be extended to all measured wavelengths and the resulting data sets are plotted as two-dimensional color maps of plasmonic interferograms (Fig.~\ref{figFFT}C) and Fourier transform power spectra (Fig.~\ref{figFFT}D), which reveal the energy-momentum dispersion of SPPs existing on both interfaces. By filtering out the higher-order SPP contributions (with $m>1$), plasmonic interferograms with only first-order SPP contributions can be reconstructed (red line in Fig.~\ref{figFFT}A), and the corresponding SPP coupling coefficients $\beta_\mathrm{t}$ and $\beta_\mathrm{b}$ can be extracted from fits to the filtered data~\cite{SuppMat}.

\begin{figure*}[h]	
\centering
\includegraphics[width=1\textwidth]{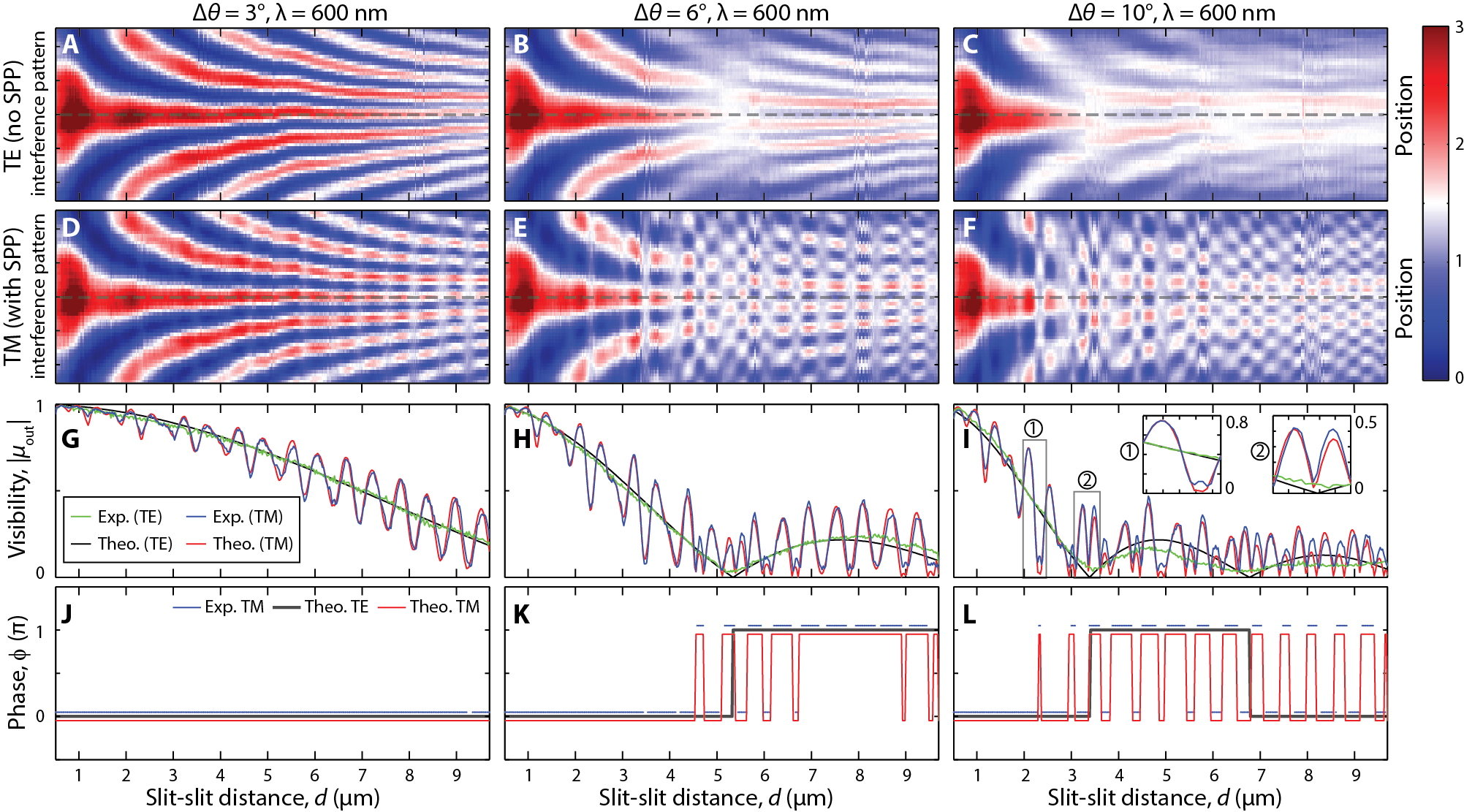}
	\caption{\textbf{SPP-enabled amplitude and phase modulation of complex degree of spatial coherence.} (\textbf{A}--\textbf{C}) TE- and (\textbf{D}--\textbf{F}) TM-polarized Young's double-slit far-zone interference patterns measured at $\lambda=600$~nm as a function of slit-slit distance, and for three different K\"ohler subtended illumination angles ($\Delta \theta \approx$~3$^\circ$, 6$^\circ$, 10$^\circ$), corresponding to decreasing spatial coherence length ($L_C=~$3.18, 1.47, and 0.93~$\mu$m, respectively). (\textbf{G}--\textbf{I}) Experimental visibility curves extracted from far-zone interference patterns measured under TE- (green lines, no SPPs) and TM-polarized (blue lines, with SPPs) illumination, at $\lambda=$~600~nm, and for different values of $\Delta \theta$. The black lines are the theoretical results for TE illumination obtained by fitting the corresponding experimental data to a \textit{sinc} function ($|\mathrm{sinc}(k_0d\Delta \theta/2)|$) with $\Delta \theta$ as the only fitting parameter. The red lines are theoretical predictions of visibility under TM illumination calculated using Eq.~(\ref{eqmu12}) and including SPPs from both metal/dielectric interfaces. \textbf{Insets in} \textbf{I} highlight strong modulation of visibility (and, correspondingly, amplitude of complex degree of spatial coherence) induced by SPPs. (\textbf{J}--\textbf{L}) Experimental TM (blue lines), theoretical TE (gray lines) and theoretical TM (red lines) phase values of complex degrees of spatial coherence for various $\Delta \theta$. For clarity, the red and blue lines are slightly shifted in the vertical direction. Discrete data points sporadically missing from blue lines correspond to visibility values $\mathcal{V}<$~0.088 for which the phase cannot be accurately retrieved.}
    \label{figVisCur}
\end{figure*}

\paragraph*{\textbf{Strong modulation of spatial coherence.}}
Far-zone interference patterns as a function of slit-slit separation distance can be constructed by scanning all of the fabricated Young's double-slit interferometers with varying separation distances under TE- (Fig.~\ref{figVisCur}, A to C) and TM-polarized (Fig.~\ref{figVisCur}, D to F) illumination, respectively. The far-zone interference patterns recorded in the presence of SPPs show strong modulation in the fringe contrast and richer intensity features, with overall increased number of observable maxima and minima (compare, for instance, Fig.~\ref{figVisCur}F with Fig.~\ref{figVisCur}C). The  amplitude and phase of the complex degree of spatial coherence at any given wavelength can be extracted from quantitative inspection of the interference patterns. The results of such an exercise are reported in Fig.~\ref{figVisCur}, G to I, and Fig.~\ref{figVisCur}, J to L, for both TE and TM  illumination, at  $\lambda=$~600~nm. The fringe visibility recorded under TE-polarized (no SPPs) illumination with $\Delta \theta=$~3$^\circ$ gradually decreases as a function of increasing slit-slit separation distance (green line in Fig.~\ref{figVisCur}G). Such a change follows the model derived from the van Cittert--Zernike theorem (black line in Fig.~\ref{figVisCur}G)~\cite{wolf2007introduction}. In contrast, the measured visibility under TM illumination (with SPPs) is significantly modulated (blue line in Fig.~\ref{figVisCur}G). It is worth noting that for small separation distances ($d<5~\mu$m), interference effects caused by beatings of SPPs supported by both metal/dielectric interfaces are clearly visible, as evidenced by the non single-periodic oscillations of the visibility curve. In contrast, for large separation distances ($d>5~\mu$m), the SPPs propagating along the glass/Ti/Ag interface are more strongly attenuated compared to the SPPs along the Ag/air interface, and no longer contribute to visibility modulation. According to Eq.~(\ref{eqmu12}), under TE  illumination $\mu_\mathrm{out}=\mu_\mathrm{in}$  since SPPs are not excited (i.e., $\beta=0$) and cannot modulate fringe visibility. Thus, the experimental visibility values extracted under TE illumination are a direct measure of the mutual correlation function of the incident light at the slit locations. From this, and by using the SPP coupling coefficients obtained from plasmonic interferograms as input parameters in Eq.~(\ref{eqmu12}), amplitude and phase of the complex  degree of spatial coherence under TM illumination can be theoretically predicted (red lines in Fig.~\ref{figVisCur}G and Fig.~\ref{figVisCur}J, respectively). The theoretical values agree well with the experimental results over the whole range of visibility, wavelength, and slit-slit separation distance~\cite{SuppMat}. 

\vspace{0.1cm}
\paragraph*{\textbf{Discussion.}}
As highlighted in the first inset to Fig.~\ref{figVisCur}I, by finely varying the slit-slit separation distance within a range of only $\sim\lambda_\mathrm{SPP}/2\approx 240$~nm, with $\lambda_\mathrm{SPP}$ the average SPP wavelength at $\lambda=$~600~nm, the experimental visibility can be varied from $\sim$0 (totally incoherent) to $\sim$0.8 (almost fully coherent) under TM illumination. Furthermore, as shown in the second inset to Fig.~\ref{figVisCur}I, by simply changing the polarization state from TE (no SPPs, green line) to TM (with SPPs, blue line), fully incoherent incident fields can be transformed into highly correlated ($\phi=0$) or anti-correlated ($\phi=\pi$) fields with $\mathcal{V}>0.4$. In addition to strong amplitude modulation of spatial coherence, SPPs on both interfaces can also be tuned to finely modulate the phase of the complex degree of spatial coherence, as shown in Fig.~\ref{figVisCur}, K and L.

In summary we have achieved full control of the complex degree of spatial coherence of light at the micrometer scale using surface-plasmon based interferometers. Strong modulation of the amplitude and phase of the complex degree of coherence is accomplished by engineering the beatings between the surface plasmons with different  wavevectors generated at the two metal/dielectric interfaces. These findings can lead to alternative nano-engineered optical flat surfaces that leverage the degree of spatial coherence to achieve ultimate control of light flow, beyond conventional refractive- and diffractive-based photonic metasurfaces.

\vspace{-0.8cm}
\begin{acknowledgments}
\vspace{-0.4cm}
This work was supported by the National Science Foundation (NSF) (grant no. CMMI–1530547).
\end{acknowledgments}

\bibliography{scibib}

\begin{thebibliography}{29}%
\makeatletter
\providecommand \@ifxundefined [1]{%
 \@ifx{#1\undefined}
}%
\providecommand \@ifnum [1]{%
 \ifnum #1\expandafter \@firstoftwo
 \else \expandafter \@secondoftwo
 \fi
}%
\providecommand \@ifx [1]{%
 \ifx #1\expandafter \@firstoftwo
 \else \expandafter \@secondoftwo
 \fi
}%
\providecommand \natexlab [1]{#1}%
\providecommand \enquote  [1]{``#1''}%
\providecommand \bibnamefont  [1]{#1}%
\providecommand \bibfnamefont [1]{#1}%
\providecommand \citenamefont [1]{#1}%
\providecommand \href@noop [0]{\@secondoftwo}%
\providecommand \href [0]{\begingroup \@sanitize@url \@href}%
\providecommand \@href[1]{\@@startlink{#1}\@@href}%
\providecommand \@@href[1]{\endgroup#1\@@endlink}%
\providecommand \@sanitize@url [0]{\catcode `\\12\catcode `\$12\catcode
  `\&12\catcode `\#12\catcode `\^12\catcode `\_12\catcode `\%12\relax}%
\providecommand \@@startlink[1]{}%
\providecommand \@@endlink[0]{}%
\providecommand \url  [0]{\begingroup\@sanitize@url \@url }%
\providecommand \@url [1]{\endgroup\@href {#1}{\urlprefix }}%
\providecommand \urlprefix  [0]{URL }%
\providecommand \Eprint [0]{\href }%
\providecommand \doibase [0]{http://dx.doi.org/}%
\providecommand \selectlanguage [0]{\@gobble}%
\providecommand \bibinfo  [0]{\@secondoftwo}%
\providecommand \bibfield  [0]{\@secondoftwo}%
\providecommand \translation [1]{[#1]}%
\providecommand \BibitemOpen [0]{}%
\providecommand \bibitemStop [0]{}%
\providecommand \bibitemNoStop [0]{.\EOS\space}%
\providecommand \EOS [0]{\spacefactor3000\relax}%
\providecommand \BibitemShut  [1]{\csname bibitem#1\endcsname}%
\let\auto@bib@innerbib\@empty
\bibitem [{\citenamefont {Mandel}\ and\ \citenamefont
  {Wolf}(1995)}]{mandel1995optical}%
  \BibitemOpen
  \bibfield  {author} {\bibinfo {author} {\bibfnamefont {L.}~\bibnamefont
  {Mandel}}\ and\ \bibinfo {author} {\bibfnamefont {E.}~\bibnamefont {Wolf}},\
  }\href@noop {} {\emph {\bibinfo {title} {Optical Coherence and Quantum
  Optics}}}\ (\bibinfo  {publisher} {Cambridge Univ. Press},\ \bibinfo {year}
  {1995})\BibitemShut {NoStop}%
\bibitem [{\citenamefont {Wolf}(2007)}]{wolf2007introduction}%
  \BibitemOpen
  \bibfield  {author} {\bibinfo {author} {\bibfnamefont {E.}~\bibnamefont
  {Wolf}},\ }\href@noop {} {\emph {\bibinfo {title} {Introduction to the Theory
  of Coherence and Polarization of Light}}}\ (\bibinfo  {publisher} {Cambridge
  University Press},\ \bibinfo {year} {2007})\BibitemShut {NoStop}%
\bibitem [{\citenamefont {Wolf}(2003)}]{Wolf_PLA}%
  \BibitemOpen
  \bibfield  {author} {\bibinfo {author} {\bibfnamefont {E.}~\bibnamefont
  {Wolf}},\ }\href@noop {} {\bibfield  {journal} {\bibinfo  {journal} {Physics
  Letters A}\ }\textbf {\bibinfo {volume} {312}},\ \bibinfo {pages} {263}
  (\bibinfo {year} {2003})}\BibitemShut {NoStop}%
\bibitem [{\citenamefont {Kildishev}\ \emph {et~al.}(2013)\citenamefont
  {Kildishev}, \citenamefont {Boltasseva},\ and\ \citenamefont
  {Shalaev}}]{kildishev2013planar}%
  \BibitemOpen
  \bibfield  {author} {\bibinfo {author} {\bibfnamefont {A.~V.}\ \bibnamefont
  {Kildishev}}, \bibinfo {author} {\bibfnamefont {A.}~\bibnamefont
  {Boltasseva}}, \ and\ \bibinfo {author} {\bibfnamefont {V.~M.}\ \bibnamefont
  {Shalaev}},\ }\href@noop {} {\bibfield  {journal} {\bibinfo  {journal}
  {Science}\ }\textbf {\bibinfo {volume} {339}},\ \bibinfo {pages} {1232009}
  (\bibinfo {year} {2013})}\BibitemShut {NoStop}%
\bibitem [{\citenamefont {Yu}\ and\ \citenamefont
  {Capasso}(2014)}]{yu2014flat}%
  \BibitemOpen
  \bibfield  {author} {\bibinfo {author} {\bibfnamefont {N.}~\bibnamefont
  {Yu}}\ and\ \bibinfo {author} {\bibfnamefont {F.}~\bibnamefont {Capasso}},\
  }\href@noop {} {\bibfield  {journal} {\bibinfo  {journal} {Nat. Materials}\
  }\textbf {\bibinfo {volume} {13}},\ \bibinfo {pages} {139} (\bibinfo {year}
  {2014})}\BibitemShut {NoStop}%
\bibitem [{\citenamefont {Lin}\ \emph {et~al.}(2014)\citenamefont {Lin},
  \citenamefont {Fan}, \citenamefont {Hasman},\ and\ \citenamefont
  {Brongersma}}]{lin2014dielectric}%
  \BibitemOpen
  \bibfield  {author} {\bibinfo {author} {\bibfnamefont {D.}~\bibnamefont
  {Lin}}, \bibinfo {author} {\bibfnamefont {P.}~\bibnamefont {Fan}}, \bibinfo
  {author} {\bibfnamefont {E.}~\bibnamefont {Hasman}}, \ and\ \bibinfo {author}
  {\bibfnamefont {M.~L.}\ \bibnamefont {Brongersma}},\ }\href@noop {}
  {\bibfield  {journal} {\bibinfo  {journal} {Science}\ }\textbf {\bibinfo
  {volume} {345}},\ \bibinfo {pages} {298} (\bibinfo {year}
  {2014})}\BibitemShut {NoStop}%
\bibitem [{\citenamefont {Yu}\ \emph {et~al.}(2011)\citenamefont {Yu},
  \citenamefont {Genevet}, \citenamefont {Kats}, \citenamefont {Aieta},
  \citenamefont {Tetienne}, \citenamefont {Capasso},\ and\ \citenamefont
  {Gaburro}}]{yu2011light}%
  \BibitemOpen
  \bibfield  {author} {\bibinfo {author} {\bibfnamefont {N.}~\bibnamefont
  {Yu}}, \bibinfo {author} {\bibfnamefont {P.}~\bibnamefont {Genevet}},
  \bibinfo {author} {\bibfnamefont {M.~A.}\ \bibnamefont {Kats}}, \bibinfo
  {author} {\bibfnamefont {F.}~\bibnamefont {Aieta}}, \bibinfo {author}
  {\bibfnamefont {J.-P.}\ \bibnamefont {Tetienne}}, \bibinfo {author}
  {\bibfnamefont {F.}~\bibnamefont {Capasso}}, \ and\ \bibinfo {author}
  {\bibfnamefont {Z.}~\bibnamefont {Gaburro}},\ }\href@noop {} {\bibfield
  {journal} {\bibinfo  {journal} {Science}\ }\textbf {\bibinfo {volume}
  {334}},\ \bibinfo {pages} {333} (\bibinfo {year} {2011})}\BibitemShut
  {NoStop}%
\bibitem [{\citenamefont {Aieta}\ \emph {et~al.}(2015)\citenamefont {Aieta},
  \citenamefont {Kats}, \citenamefont {Genevet},\ and\ \citenamefont
  {Capasso}}]{aieta2015multiwavelength}%
  \BibitemOpen
  \bibfield  {author} {\bibinfo {author} {\bibfnamefont {F.}~\bibnamefont
  {Aieta}}, \bibinfo {author} {\bibfnamefont {M.~A.}\ \bibnamefont {Kats}},
  \bibinfo {author} {\bibfnamefont {P.}~\bibnamefont {Genevet}}, \ and\
  \bibinfo {author} {\bibfnamefont {F.}~\bibnamefont {Capasso}},\ }\href@noop
  {} {\bibfield  {journal} {\bibinfo  {journal} {Science}\ }\textbf {\bibinfo
  {volume} {347}},\ \bibinfo {pages} {1342} (\bibinfo {year}
  {2015})}\BibitemShut {NoStop}%
\bibitem [{\citenamefont {Maguid}\ \emph {et~al.}(2016)\citenamefont {Maguid},
  \citenamefont {Yulevich}, \citenamefont {Veksler}, \citenamefont {Kleiner},
  \citenamefont {Brongersma},\ and\ \citenamefont
  {Hasman}}]{maguid2016photonic}%
  \BibitemOpen
  \bibfield  {author} {\bibinfo {author} {\bibfnamefont {E.}~\bibnamefont
  {Maguid}}, \bibinfo {author} {\bibfnamefont {I.}~\bibnamefont {Yulevich}},
  \bibinfo {author} {\bibfnamefont {D.}~\bibnamefont {Veksler}}, \bibinfo
  {author} {\bibfnamefont {V.}~\bibnamefont {Kleiner}}, \bibinfo {author}
  {\bibfnamefont {M.~L.}\ \bibnamefont {Brongersma}}, \ and\ \bibinfo {author}
  {\bibfnamefont {E.}~\bibnamefont {Hasman}},\ }\href@noop {} {\bibfield
  {journal} {\bibinfo  {journal} {Science}\ }\textbf {\bibinfo {volume}
  {352}},\ \bibinfo {pages} {1202} (\bibinfo {year} {2016})}\BibitemShut
  {NoStop}%
\bibitem [{\citenamefont {Redding}\ \emph {et~al.}(2012)\citenamefont
  {Redding}, \citenamefont {Choma},\ and\ \citenamefont
  {Cao}}]{Redding_NatPhoton}%
  \BibitemOpen
  \bibfield  {author} {\bibinfo {author} {\bibfnamefont {B.}~\bibnamefont
  {Redding}}, \bibinfo {author} {\bibfnamefont {M.~A.}\ \bibnamefont {Choma}},
  \ and\ \bibinfo {author} {\bibfnamefont {H.}~\bibnamefont {Cao}},\
  }\href@noop {} {\bibfield  {journal} {\bibinfo  {journal} {Nat. Photonics}\
  }\textbf {\bibinfo {volume} {6}},\ \bibinfo {pages} {355} (\bibinfo {year}
  {2012})}\BibitemShut {NoStop}%
\bibitem [{\citenamefont {Knitter}\ \emph {et~al.}(2016)\citenamefont
  {Knitter}, \citenamefont {Liu}, \citenamefont {Redding}, \citenamefont
  {Khokha}, \citenamefont {Choma},\ and\ \citenamefont {Cao}}]{Knitter}%
  \BibitemOpen
  \bibfield  {author} {\bibinfo {author} {\bibfnamefont {S.}~\bibnamefont
  {Knitter}}, \bibinfo {author} {\bibfnamefont {C.}~\bibnamefont {Liu}},
  \bibinfo {author} {\bibfnamefont {B.}~\bibnamefont {Redding}}, \bibinfo
  {author} {\bibfnamefont {M.~K.}\ \bibnamefont {Khokha}}, \bibinfo {author}
  {\bibfnamefont {M.~A.}\ \bibnamefont {Choma}}, \ and\ \bibinfo {author}
  {\bibfnamefont {H.}~\bibnamefont {Cao}},\ }\href@noop {} {\bibfield
  {journal} {\bibinfo  {journal} {Optica}\ }\textbf {\bibinfo {volume} {3}},\
  \bibinfo {pages} {403} (\bibinfo {year} {2016})}\BibitemShut {NoStop}%
\bibitem [{\citenamefont {Kip}\ \emph {et~al.}(2000)\citenamefont {Kip},
  \citenamefont {Soljacic}, \citenamefont {Segev}, \citenamefont {Eugenieva},\
  and\ \citenamefont {Christodoulides}}]{kip2000modulation}%
  \BibitemOpen
  \bibfield  {author} {\bibinfo {author} {\bibfnamefont {D.}~\bibnamefont
  {Kip}}, \bibinfo {author} {\bibfnamefont {M.}~\bibnamefont {Soljacic}},
  \bibinfo {author} {\bibfnamefont {M.}~\bibnamefont {Segev}}, \bibinfo
  {author} {\bibfnamefont {E.}~\bibnamefont {Eugenieva}}, \ and\ \bibinfo
  {author} {\bibfnamefont {D.~N.}\ \bibnamefont {Christodoulides}},\
  }\href@noop {} {\bibfield  {journal} {\bibinfo  {journal} {Science}\ }\textbf
  {\bibinfo {volume} {290}},\ \bibinfo {pages} {495} (\bibinfo {year}
  {2000})}\BibitemShut {NoStop}%
\bibitem [{\citenamefont {van Dijk}\ \emph {et~al.}(2010)\citenamefont {van
  Dijk}, \citenamefont {Fischer}, \citenamefont {Visser},\ and\ \citenamefont
  {Wolf}}]{Dijk_PRL}%
  \BibitemOpen
  \bibfield  {author} {\bibinfo {author} {\bibfnamefont {T.}~\bibnamefont {van
  Dijk}}, \bibinfo {author} {\bibfnamefont {D.~G.}\ \bibnamefont {Fischer}},
  \bibinfo {author} {\bibfnamefont {T.~D.}\ \bibnamefont {Visser}}, \ and\
  \bibinfo {author} {\bibfnamefont {E.}~\bibnamefont {Wolf}},\ }\href@noop {}
  {\bibfield  {journal} {\bibinfo  {journal} {Phys. Rev. Lett.}\ }\textbf
  {\bibinfo {volume} {104}},\ \bibinfo {pages} {173902} (\bibinfo {year}
  {2010})}\BibitemShut {NoStop}%
\bibitem [{\citenamefont {Wang}\ \emph {et~al.}(2015)\citenamefont {Wang},
  \citenamefont {Yan}, \citenamefont {Kuebel},\ and\ \citenamefont
  {Visser}}]{Wang_PRA_2015}%
  \BibitemOpen
  \bibfield  {author} {\bibinfo {author} {\bibfnamefont {Y.}~\bibnamefont
  {Wang}}, \bibinfo {author} {\bibfnamefont {S.}~\bibnamefont {Yan}}, \bibinfo
  {author} {\bibfnamefont {D.}~\bibnamefont {Kuebel}}, \ and\ \bibinfo {author}
  {\bibfnamefont {T.~D.}\ \bibnamefont {Visser}},\ }\href@noop {} {\bibfield
  {journal} {\bibinfo  {journal} {Phys. Rev. A}\ }\textbf {\bibinfo {volume}
  {92}},\ \bibinfo {pages} {013806} (\bibinfo {year} {2015})}\BibitemShut
  {NoStop}%
\bibitem [{\citenamefont {Wang}\ \emph {et~al.}(2016)\citenamefont {Wang},
  \citenamefont {Kuebel}, \citenamefont {Visser},\ and\ \citenamefont
  {Wolf}}]{Wang_PRA_2016}%
  \BibitemOpen
  \bibfield  {author} {\bibinfo {author} {\bibfnamefont {Y.}~\bibnamefont
  {Wang}}, \bibinfo {author} {\bibfnamefont {D.}~\bibnamefont {Kuebel}},
  \bibinfo {author} {\bibfnamefont {T.~D.}\ \bibnamefont {Visser}}, \ and\
  \bibinfo {author} {\bibfnamefont {E.}~\bibnamefont {Wolf}},\ }\href@noop {}
  {\bibfield  {journal} {\bibinfo  {journal} {Phys. Rev. A}\ }\textbf {\bibinfo
  {volume} {94}},\ \bibinfo {pages} {033812} (\bibinfo {year}
  {2016})}\BibitemShut {NoStop}%
\bibitem [{\citenamefont {Anderson}\ and\ \citenamefont
  {Pelz}(1995)}]{anderson1995spatial}%
  \BibitemOpen
  \bibfield  {author} {\bibinfo {author} {\bibfnamefont {B.~L.}\ \bibnamefont
  {Anderson}}\ and\ \bibinfo {author} {\bibfnamefont {L.~J.}\ \bibnamefont
  {Pelz}},\ }\href@noop {} {\bibfield  {journal} {\bibinfo  {journal} {Applied
  Optics}\ }\textbf {\bibinfo {volume} {34}},\ \bibinfo {pages} {7443}
  (\bibinfo {year} {1995})}\BibitemShut {NoStop}%
\bibitem [{\citenamefont {Gbur}(2014)}]{Gbur_JOSAA}%
  \BibitemOpen
  \bibfield  {author} {\bibinfo {author} {\bibfnamefont {G.}~\bibnamefont
  {Gbur}},\ }\href {\doibase 10.1364/JOSAA.31.002038} {\bibfield  {journal}
  {\bibinfo  {journal} {J. Opt. Soc. Am. A}\ }\textbf {\bibinfo {volume}
  {31}},\ \bibinfo {pages} {2038} (\bibinfo {year} {2014})}\BibitemShut
  {NoStop}%
\bibitem [{\citenamefont {Gan}\ \emph {et~al.}(2007)\citenamefont {Gan},
  \citenamefont {Gbur},\ and\ \citenamefont {Visser}}]{Gan_PRL}%
  \BibitemOpen
  \bibfield  {author} {\bibinfo {author} {\bibfnamefont {C.~H.}\ \bibnamefont
  {Gan}}, \bibinfo {author} {\bibfnamefont {G.}~\bibnamefont {Gbur}}, \ and\
  \bibinfo {author} {\bibfnamefont {T.~D.}\ \bibnamefont {Visser}},\ }\href
  {\doibase 10.1103/PhysRevLett.98.043908} {\bibfield  {journal} {\bibinfo
  {journal} {Phys. Rev. Lett.}\ }\textbf {\bibinfo {volume} {98}},\ \bibinfo
  {pages} {043908} (\bibinfo {year} {2007})}\BibitemShut {NoStop}%
\bibitem [{\citenamefont {Gan}\ \emph {et~al.}(2012)\citenamefont {Gan},
  \citenamefont {Gu}, \citenamefont {Visser},\ and\ \citenamefont
  {Gbur}}]{Gan_Plasm}%
  \BibitemOpen
  \bibfield  {author} {\bibinfo {author} {\bibfnamefont {C.~H.}\ \bibnamefont
  {Gan}}, \bibinfo {author} {\bibfnamefont {Y.}~\bibnamefont {Gu}}, \bibinfo
  {author} {\bibfnamefont {T.~D.}\ \bibnamefont {Visser}}, \ and\ \bibinfo
  {author} {\bibfnamefont {G.}~\bibnamefont {Gbur}},\ }\href@noop {} {\bibfield
   {journal} {\bibinfo  {journal} {Plasmonics}\ }\textbf {\bibinfo {volume}
  {7}},\ \bibinfo {pages} {313} (\bibinfo {year} {2012})}\BibitemShut {NoStop}%
\bibitem [{\citenamefont {Kuzmin}\ \emph {et~al.}(2007)\citenamefont {Kuzmin},
  \citenamefont {Hooft}, \citenamefont {Eliel}, \citenamefont {Gbur},
  \citenamefont {Schouten},\ and\ \citenamefont {Visser}}]{Kuzmin}%
  \BibitemOpen
  \bibfield  {author} {\bibinfo {author} {\bibfnamefont {N.}~\bibnamefont
  {Kuzmin}}, \bibinfo {author} {\bibfnamefont {G.}~\bibnamefont {Hooft}},
  \bibinfo {author} {\bibfnamefont {E.}~\bibnamefont {Eliel}}, \bibinfo
  {author} {\bibfnamefont {G.}~\bibnamefont {Gbur}}, \bibinfo {author}
  {\bibfnamefont {H.}~\bibnamefont {Schouten}}, \ and\ \bibinfo {author}
  {\bibfnamefont {T.}~\bibnamefont {Visser}},\ }\href@noop {} {\bibfield
  {journal} {\bibinfo  {journal} {Optics Lett.}\ }\textbf {\bibinfo {volume}
  {32}},\ \bibinfo {pages} {445} (\bibinfo {year} {2007})}\BibitemShut
  {NoStop}%
\bibitem [{\citenamefont {Ravets}\ \emph {et~al.}(2009)\citenamefont {Ravets},
  \citenamefont {Rodier}, \citenamefont {Kim}, \citenamefont {Hugonin},
  \citenamefont {Jacubowiez},\ and\ \citenamefont {Lalanne}}]{Ravets}%
  \BibitemOpen
  \bibfield  {author} {\bibinfo {author} {\bibfnamefont {S.}~\bibnamefont
  {Ravets}}, \bibinfo {author} {\bibfnamefont {J.-C.}\ \bibnamefont {Rodier}},
  \bibinfo {author} {\bibfnamefont {B.~E.}\ \bibnamefont {Kim}}, \bibinfo
  {author} {\bibfnamefont {J.-P.}\ \bibnamefont {Hugonin}}, \bibinfo {author}
  {\bibfnamefont {L.}~\bibnamefont {Jacubowiez}}, \ and\ \bibinfo {author}
  {\bibfnamefont {P.}~\bibnamefont {Lalanne}},\ }\href@noop {} {\bibfield
  {journal} {\bibinfo  {journal} {JOSA B}\ }\textbf {\bibinfo {volume} {26}},\
  \bibinfo {pages} {B28} (\bibinfo {year} {2009})}\BibitemShut {NoStop}%
\bibitem [{\citenamefont {Divitt}\ \emph {et~al.}(2016)\citenamefont {Divitt},
  \citenamefont {Frimmer}, \citenamefont {Visser},\ and\ \citenamefont
  {Novotny}}]{Divitt_OpLett}%
  \BibitemOpen
  \bibfield  {author} {\bibinfo {author} {\bibfnamefont {S.}~\bibnamefont
  {Divitt}}, \bibinfo {author} {\bibfnamefont {M.}~\bibnamefont {Frimmer}},
  \bibinfo {author} {\bibfnamefont {T.~D.}\ \bibnamefont {Visser}}, \ and\
  \bibinfo {author} {\bibfnamefont {L.}~\bibnamefont {Novotny}},\ }\href@noop
  {} {\bibfield  {journal} {\bibinfo  {journal} {Optics Lett.}\ }\textbf
  {\bibinfo {volume} {41}},\ \bibinfo {pages} {3094} (\bibinfo {year}
  {2016})}\BibitemShut {NoStop}%
\bibitem [{\citenamefont {Mandel}\ and\ \citenamefont
  {Wolf}(1976)}]{Mandel_JOSA}%
  \BibitemOpen
  \bibfield  {author} {\bibinfo {author} {\bibfnamefont {L.}~\bibnamefont
  {Mandel}}\ and\ \bibinfo {author} {\bibfnamefont {E.}~\bibnamefont {Wolf}},\
  }\href@noop {} {\bibfield  {journal} {\bibinfo  {journal} {JOSA}\ }\textbf
  {\bibinfo {volume} {66}},\ \bibinfo {pages} {529} (\bibinfo {year}
  {1976})}\BibitemShut {NoStop}%
\bibitem [{\citenamefont {Altewischer}\ \emph {et~al.}(2002)\citenamefont
  {Altewischer}, \citenamefont {Van~Exter},\ and\ \citenamefont
  {Woerdman}}]{Altewischer}%
  \BibitemOpen
  \bibfield  {author} {\bibinfo {author} {\bibfnamefont {E.}~\bibnamefont
  {Altewischer}}, \bibinfo {author} {\bibfnamefont {M.}~\bibnamefont
  {Van~Exter}}, \ and\ \bibinfo {author} {\bibfnamefont {J.}~\bibnamefont
  {Woerdman}},\ }\href@noop {} {\bibfield  {journal} {\bibinfo  {journal}
  {Nature}\ }\textbf {\bibinfo {volume} {418}},\ \bibinfo {pages} {304}
  (\bibinfo {year} {2002})}\BibitemShut {NoStop}%
\bibitem [{\citenamefont {Tame}\ \emph {et~al.}(2013)\citenamefont {Tame},
  \citenamefont {McEnery}, \citenamefont {{\"O}zdemir}, \citenamefont {Lee},
  \citenamefont {Maier},\ and\ \citenamefont {Kim}}]{Tame}%
  \BibitemOpen
  \bibfield  {author} {\bibinfo {author} {\bibfnamefont {M.~S.}\ \bibnamefont
  {Tame}}, \bibinfo {author} {\bibfnamefont {K.}~\bibnamefont {McEnery}},
  \bibinfo {author} {\bibfnamefont {{\c{S}}.}~\bibnamefont {{\"O}zdemir}},
  \bibinfo {author} {\bibfnamefont {J.}~\bibnamefont {Lee}}, \bibinfo {author}
  {\bibfnamefont {S.}~\bibnamefont {Maier}}, \ and\ \bibinfo {author}
  {\bibfnamefont {M.}~\bibnamefont {Kim}},\ }\href@noop {} {\bibfield
  {journal} {\bibinfo  {journal} {Nat. Phys.}\ }\textbf {\bibinfo {volume}
  {9}},\ \bibinfo {pages} {329} (\bibinfo {year} {2013})}\BibitemShut {NoStop}%
\bibitem [{\citenamefont {Wang}\ \emph {et~al.}(2014)\citenamefont {Wang},
  \citenamefont {Comtet}, \citenamefont {Le~Moal}, \citenamefont {Dujardin},
  \citenamefont {Drezet}, \citenamefont {Huant},\ and\ \citenamefont
  {Boer-Duchemin}}]{Wang_OpLett}%
  \BibitemOpen
  \bibfield  {author} {\bibinfo {author} {\bibfnamefont {T.}~\bibnamefont
  {Wang}}, \bibinfo {author} {\bibfnamefont {G.}~\bibnamefont {Comtet}},
  \bibinfo {author} {\bibfnamefont {E.}~\bibnamefont {Le~Moal}}, \bibinfo
  {author} {\bibfnamefont {G.}~\bibnamefont {Dujardin}}, \bibinfo {author}
  {\bibfnamefont {A.}~\bibnamefont {Drezet}}, \bibinfo {author} {\bibfnamefont
  {S.}~\bibnamefont {Huant}}, \ and\ \bibinfo {author} {\bibfnamefont
  {E.}~\bibnamefont {Boer-Duchemin}},\ }\href@noop {} {\bibfield  {journal}
  {\bibinfo  {journal} {Optics Lett.}\ }\textbf {\bibinfo {volume} {39}},\
  \bibinfo {pages} {6679} (\bibinfo {year} {2014})}\BibitemShut {NoStop}%
\bibitem [{Sup()}]{SuppMat}%
  \BibitemOpen
  \href@noop {} {\bibinfo  {journal} {See Supplementary Materials for details}\
  }\BibitemShut {NoStop}%
\bibitem [{\citenamefont {Morrill}\ \emph {et~al.}(2016)\citenamefont
  {Morrill}, \citenamefont {Li},\ and\ \citenamefont {Pacifici}}]{Morrill}%
  \BibitemOpen
\bibfield  {journal} {  }\bibfield  {author} {\bibinfo {author} {\bibfnamefont
  {D.}~\bibnamefont {Morrill}}, \bibinfo {author} {\bibfnamefont
  {D.}~\bibnamefont {Li}}, \ and\ \bibinfo {author} {\bibfnamefont
  {D.}~\bibnamefont {Pacifici}},\ }\href@noop {} {\bibfield  {journal}
  {\bibinfo  {journal} {Nat. Photonics}\ }\textbf {\bibinfo {volume} {10}},\
  \bibinfo {pages} {681} (\bibinfo {year} {2016})}\BibitemShut {NoStop}%
\bibitem [{\citenamefont {Li}\ \emph {et~al.}(2016)\citenamefont {Li},
  \citenamefont {Feng},\ and\ \citenamefont {Pacifici}}]{Li_OE}%
  \BibitemOpen
  \bibfield  {author} {\bibinfo {author} {\bibfnamefont {D.}~\bibnamefont
  {Li}}, \bibinfo {author} {\bibfnamefont {J.}~\bibnamefont {Feng}}, \ and\
  \bibinfo {author} {\bibfnamefont {D.}~\bibnamefont {Pacifici}},\ }\href@noop
  {} {\bibfield  {journal} {\bibinfo  {journal} {Opt. Express}\ }\textbf
  {\bibinfo {volume} {24}},\ \bibinfo {pages} {27309} (\bibinfo {year}
  {2016})}\BibitemShut {NoStop}%
\end{thebibliography}%


\end{document}